\documentclass[a4paper, 11pt]{article}
\usepackage[T1]{fontenc} % if needed
\usepackage[utf8]{inputenc}
\usepackage[english]{babel}
\usepackage{authblk}
\usepackage{graphicx}%
\usepackage{multirow}%
\usepackage{amsmath,amssymb,amsfonts}%
\usepackage{amsthm}%
\usepackage{hyperref}
\usepackage{mathrsfs}%
\usepackage[title]{appendix}%
\usepackage{xcolor}%
\usepackage{textcomp}%
\usepackage{manyfoot}%
\usepackage{booktabs}%
\usepackage{algorithm}%
\usepackage{algorithmicx}%
\usepackage{algpseudocode}%
\usepackage{listings}%
\usepackage{siunitx}
\usepackage[export]{adjustbox}
\usepackage{bm}
\usepackage{csquotes}
\usepackage{adjustbox}
\DeclareSIUnit\cluce{c}

\usepackage[sorting=none,citestyle=numeric-comp, natbib]{biblatex}
\bibliography{biblio}

\graphicspath{{images/}}

\title{Baseline filtering and peak reconstruction for haloscope-like axion searches}

\author[1]{Andrea \textsc{Gallo Rosso}\thanks{\href{mailto:andrea.gallo.rosso@fysik.su.se}
{andrea.gallo.rosso@fysik.su.se}.}}
\author[1]{Jan \textsc{Conrad}}
\author[1]{Junu \textsc{Jeong}}
\affil[1]{Physics Department and Oskar Klein Centre, 
Stockholm University, Stockholm, Sweden}

\begin{document}

\maketitle

\begin{abstract}
        Axions are well-motivated dark matter 
        particles. Many experiments are looking for 
        their experimental evidence. For 
        haloscopes, the problem reduces to the 
        identification of a peak  above a noisy 
        baseline. Its modeling, however, may 
        problematic. State-of-the-art analysis rely 
        on the Savitzky-Golay (SG) filtering, which 
        is intrinsically affected by any possible 
        over fluctuation, leading to biased 
        results. In this paper we study the 
        efficiency that different extensions of SG 
        can provide in the peak reconstruction in a 
        standard haloscope-like experiment. We show 
        that, once the correlations among bins are 
        taken into account, there is no appreciable 
        difference. The standard SG remains the 
        advisable choice because of its numerical 
        efficiency.
    \end{abstract}

%%%%%%%%%%%%%%%%%%%%%%%%%%%%%%%%%%%%%%%%%%%%%%%%%%%%
\section{Introduction}\label{sec1}
%%%%%%%%%%%%%%%%%%%%%%%%%%%%%%%%%%%%%%%%%%%%%%%%%%%%
Strong evidences in astronomy and cosmology
\cite{Rubin:1970zza,Sofue:2000jx,%
Peebles:1982ff,Springel:2005nw,Clowe:2006eq,%
WMAP:2008lyn,Planck:2015fie}
suggest that about 85\% of the matter in our
universe does not belong to the Standard Model
of particle physics. The question around its 
fundamental nature is one of the most important
open problem in physics, with many candidates
being proposed theoretically---see e.g.
\cite{Bertone:2004pz,WIMPBook2017,%
Bertone:2018krk,Green:2022hhj}.
Among them, axions hold as one of the best
motivated. They were proposed as
pseudoscalar particles solving the strong CP 
problem \cite{Smith:1957ht,Dress:1976bq,%
tHooft:1976rip,tHooft:1976snw}, initially by
Peccei and Quinn \cite{Peccei:1977hh,%
Peccei:1977ur}, and then by Weinberg
\cite{Weinberg:1977ma} and Wilczek
\cite{Wilczek:1977pj}. A few years later,
it was realized that axions have the right
properties to explain dark matter
\cite{Preskill:1982cy,Abbott:1982af,Dine:1982ah}.
Axions are spin-0, stable, massive particles.
They interact weakly with the particles of the
Standard Model, and can be produced (cold and with
the right abundance) in the early Universe.

The goal to detect axions in a laboratory is
driving a titanic experimental effort
\cite{Irastorza:2018dyq,Billard:2021uyg,%
Irastorza:2021tdu,Semertzidis:2021rxs,%
Adams:2855525,AxioBook2022}, with the current
goal set by the theoretical benchmarks of
Kim-Shifman-Vainshtein-Zakharov (KSVZ) 
\cite{Kim:1979if,Shifman:1979if} and
Dine-Fischler-Srednicki-Zhitnitskii (DFSZ)
\cite{Dine:1981rt,Zhitnitsky:1980tq} models.

At present, the most promising technology to
detect axions try to measure their coupling with 
photons. Resonant cavities, or haloscopes
\cite{Sikivie:1983ip,Sikivie:1985yu} have proven
themselves able to reach the sensitivity goal for
masses below $\sim\SI{1}{\electronvolt\per%
\cluce\squared}$ \cite{ADMX:2018gho,%
ADMX:2019uok,ADMX:2021nhd,Yi:2022fmn}. For
such experiments, the sensitivity is enhanced by
the resonant conversion of axions to photons,
appearing as a narrow peak on top of the cavity
profile (in the frequency domain).
In many experiments, one of the main tasks that 
the analysis faces is to separate the 
possible signal from the known noise, in a
problem of background subtraction \cite{%
Pandey:2024dcd,%    ADBC
ADMX:2020hay,%      ADMX
Yi:2022fmn,%        CAPP
Oshima:2023csb,%    DANCE
Grenet:2021vbb,%    GraHal
Palken:2020wgs,%    HAYSTACK
Quiskamp:2023ehr,%  ORGAN
QUAX:2023gop,%      QUAX
Ahyoune:2024klt,%   RADES
Gramolin:2020ict,%  SHAFT
TASEH:2022vvu,%     TASEH
Thomson:2023moc%    UPLOAD/DOWNLOAD
}.

In principle, the best possible approach would
require a perfect knowledge of the background,
which is in principle required by any likelihood
approach (see e.g.\ \cite{Foster:2017hbq}). In
practice, this is not always possible. The most
common procedure, based on \cite{Palken:2020wgs},
is to fit the the output of the DAQ pipeline
(baseline) with a polynomial of a given degree,
performed in a sliding window over the raw data.
This is known as Savitzky-Golay (SG) filter
\cite{SavGolf}. Not only such an approach is free
from background modeling, but it is also
numerically efficient and its estimates are
analytical and predictable. The latter are
advantages that often outweigh the known issues,
such as biases, numerical artifacts and
underestimation of possible signals
\cite{Schmid:2022, Yi:2023ekw}.

In this paper, we will assess the underestimation
that an axion-looking experiment may face, in the
presence of a signal. We will make use of the
definition \cite{Yi:2023ekw} of peak reconstruction,
as well as Monte Carlo estimation. We apply the
calculation to the standard  SG 
algorithm and to possible extension thought of as
to circumvent the issue. Those methods are defined
and discussed in Section \ref{sec:baseRec}. Section
\ref{sec:effRec} presents the discussion about the
quantification of the reconstructed efficiencies.
Section \ref{sec:conc} draws the conclusions.

%%%%%%%%%%%%%%%%%%%%%%%%%%%%%%%%%%%%%%%%%%%%%%%%%%%%
\section{Formalism and method}
\label{sec:baseRec}
%%%%%%%%%%%%%%%%%%%%%%%%%%%%%%%%%%%%%%%%%%%%%%%%%%%%
In the context of axion search, standard data are
obtained by measuring the electromagnetic power
coming out of a resonant cavity, in a setup similar
e.g.\ to \cite{Brubaker:2017rna,Brubaker:2017ohw}.
In such a setup, the power coming from the
cavity are initially collected in the frequency
domain, discretized with a bandwidth resolution
$\Delta\nu$. Typical values are
$\Delta\nu\sim\SI{100}{\hertz}$. On a large
scale, in the absence of any axion signal, the
thermal noise from a cavity follows
a baseline $L(\nu_i)\equiv L_i = b_i$
which, ideally, is expected to be Lorentzian.
However, the non-ideal nature of the amplification
chain introduces in the component $b_i$ features
that are strongly dependent on the conditions of
the apparatus and might be difficult to model.

On top of that, any axions depositing power in
the cavity would appear as an excess $s_i$ 
distributed over many bins, following the profile
of a boosted Maxwellian \cite{Turner:1990qx}. In
the $i$-th bin corresponding to the frequency $
\nu_i$, the power signal $s_i$ generated by the
axion field can be written as:
\begin{equation}\label{eq:power}
    s(\nu_i)= s_i= P_{a}\int_{\nu_a +i\,\Delta\nu}
    ^{\nu_a+(i+1)\,\Delta\nu}f_a(\nu)\mathrm{d}\nu
    \approx P_a\,\Delta\nu\,f_a(\nu_i),
\end{equation}
where $P_a$ is the total deposited power and
$f_a(\nu)$ is normalized. Equation~\eqref{eq:power}
applies when the bandwidth of the cavity is much
greater than the axion linewidth.

The outcome of the experiment is constituted by 
the baseline $L_i$, which is composed by the
component $b_i$, considered as a slow-varying 
background, and possibly by the signal $s_i$. That 
is, $L_i = b_i + s_i$. In turn, 
each collected bin $[\nu_i,\,\nu_i+\Delta\nu]$
represent a fluctuation of the noise power around
the baseline, which is Gaussian with good
approximation \cite{Brubaker:2017rna,%
Palken:2020wgs}. This, assuming that the
averaging process of the acquired spectra are
sufficient.

In this paper, we focus on the extraction of
the signal component $s_i$, when only the 
combination $L_i=b_i+s_i$ is known through the
data $y(\nu_i) \equiv y_i$, and no
description of $b_i$ is available. This is usually
the case for haloscopes and  the fist step in
their analyses \cite{Brubaker:2017rna,%
Palken:2020wgs}. In fact, they focus
on the study of normalized fluctuations with
respect to the baseline, which needs to be known.
In other terms, we may consider
\begin{equation}
    y_i\sim\text{Gaus}(L_i,\, \sigma),
\end{equation}
where $\sigma$ is the fluctuation of the power noise.
The goal is to get a bin-to-bin
estimate $\hat{b}_i$ from the combination $L_i=b_i+
s_i$ provided by the data $y_i$. %\eqref{eq:def:yi}.
Then, the estimated signal $\hat{s}_i$ can be 
reconstructed as
\begin{equation}\label{eq:hatSi}
    \hat{s}_i = y_i - \hat{b}_i.
\end{equation}

The standard procedure for the estimation of 
$\hat{b}_i$ relies on the fact that $b_i$ is
smooth enough to be locally approximate as a
polynomial, even on scales of hundreds of bins.
Moreover, the scale of the variations is small
with respect to the cavity profile. Then, one
easy way to get $\hat{b}_i$ would consist in 
clearing the data from any small-scale variations.
In other word, ``smoothing-out'' $L_i$ to get
$\hat{b}_i\approx\hat{L}_i$.

The most common method to reach the goal is the
SG filtering \cite{SavGolf}. It is
characterized by numerical efficiency, linearity
and predictability of the results. In fact,
SG filtering works by means of a
polynomial fit of degree $n$ on the $N_p$ data
points in the vicinity of the $i$-th bin. For
instance, in the case of a symmetric window of
half-length $M$ around $i$, the windows becomes
$\bm{x}_i = (i-M,\,i-M+1,\dots,\,i,\,\dots,\,i+M)$,
corresponding to a same number of measured values
$\bm{y}_i=(y_{i-M},\,y_{i-M+1},\dots,\,y_{i},\,
\dots,\,y_{i+M})$. The output for the $i$-th data
point corresponds to the fitting function $f_i(x)$
evaluated at $x=i$. Because the problem is linear
(see \cite{Schafer2011,NumRec2,NumRec3} and 
Appendix \ref{App:SavGol}) the output can be
thought of as an averaged mean completely defined
by a set of  coefficients $\{c_k\}$:
\begin{equation}
    \label{eq:defSGen}
    f_i = \sum_{\{\ell\}} c_k\,y_{i+\ell}\equiv
    \hat{b}_{i}^{\rm SG}.
\end{equation}

\subsection{Extension of SG filtering}
\label{sec:FiltMeth}

\begin{figure}[t]
    \centering
    \includegraphics[width=0.75\textwidth,%
    valign=t]%
    {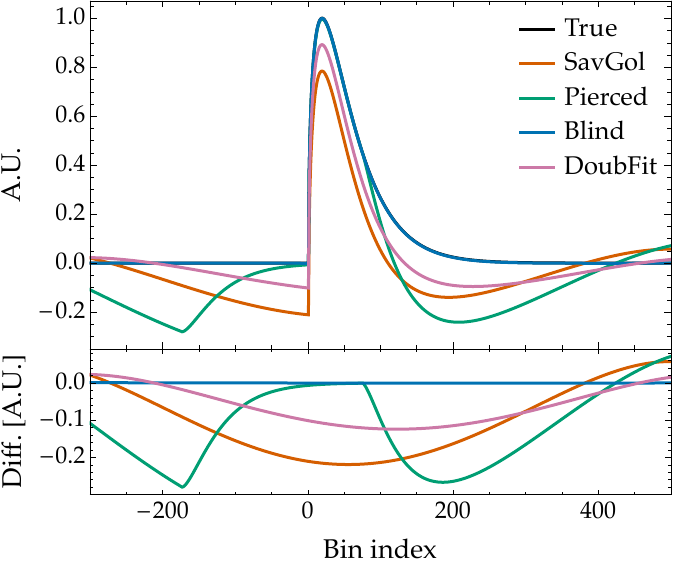}%
	\caption{Reconstructed signal $\hat{s}_i $
    \protect\eqref{eq:hatSi} from the
    (un-fluctuated) baseline $b_i +s_i$, where
    $b_i \equiv 0$ and $s_i$ parameterized by 
    equation \eqref{eq:defSi}. The right panel
    gives a graphical representation of the
    different filtering methods.}
	\label{fig:ALor}
\end{figure}

Despite its numerical efficiency, SG 
filtering is intrinsically affected by the 
presence of any signal. This happens because the
combination $L_i=b_i+s_i$ is used for an 
estimation for $b_i$  only, which causes the
results to be biased. As a consequence, the
magnitude of $s_i$ may get underestimated.
This is shown graphically in
Figure \ref{fig:ALor} (orange). There, the
SG filter is applied to a peak rising
from a flat baseline. Expressing the peak with
respect to the filtered baseline return a result 
that is lower; i.e.\, the peak is shifted downwards.

This section presents possible extension of
SG filtering aimed at overcoming this
issue. After the methods are presented, they will
be quantified in terms of their efficiency in
recovering the signal. We will make
use  both of the standard definition of efficiency
(see e.g.\ \cite{Yi:2023ekw}) and a Monte
Carlo-based analysis based on a full likelihood
approach \cite{GalloRosso:2022mhx}.

\begin{figure}[t]
    \centering
    \includegraphics[width=\textwidth,%
    valign=t]%
    {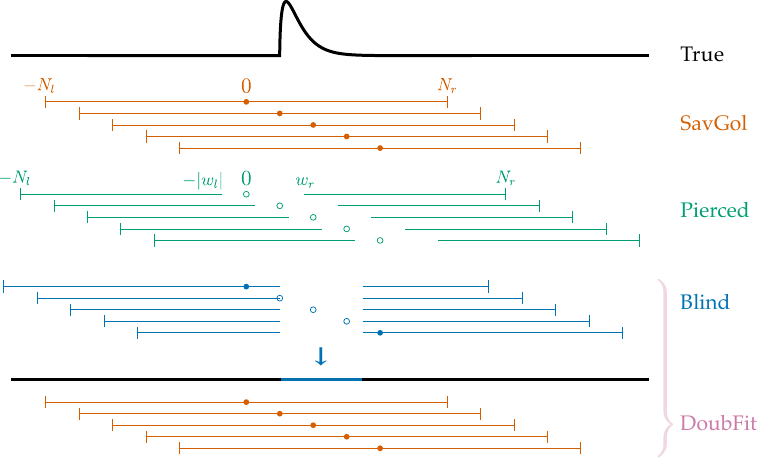}%
	\caption{Graphical representation of the
    different filtering methods.}
	\label{fig:Legend}
\end{figure}

The first filtering algorithm will be denoted as
``SavGol'' and will correspond to a classic
SG filtering over a symmetric window
of length $N_p=N_l+N_r+1$, where $N_l$ (and $N_r$)
are the number of points to the left (to the right)
of the $i$-th filtered point. Therefore, this
algorithm is fully defined by the values of the
polynomial order $n$ and the number of points
$N_p$.
With respect to the former, the results presented
in this paper are very mildly dependent on its
exact value, provided that it is high enough $(n
\gtrapprox 2)$. This is because of the smooth profile of the signal. In the following, we will set
$n = 4$.

The number of points $N_p$ needs to be sufficiently 
large to smooth out any possible signal peak from
the baseline, but also small enough to preserve its
features. In the following, it has been chosen to 
be $1201$. This is a typical value for these kind 
of analysis on real data. The results remains very
similar even for other values of $N_p$, provided
they satisfy the conditions above. The problem of
finding the optimal window length will be discussed 
in a following paper (in preparation).

The first extension of the standard  SG 
filtering algorithm is aimed at getting rid of the 
biased data. We can achieve this goal by piercing
the window around the $i$-th reconstructed point.
In this way, the baseline for the bins belonging to
the  peak is reconstructed from data points that are
away from the peak. In this case, the data that
contribute to $\hat{L}_i$ are such that $s_i\approx
0$. As a result, the reconstructed $\hat{s}_i$
would be higher, and the bias would move to the
tails. This scenario will be denoted as 
``Pierced.'' In this case, the filtering window
is defined by 4 parameters. In addition to
$N_l$ and $N_r$, two parameters define the hole 
around the filtered point, made up by $w_l>0$ 
ignored points on the left-hand side and $w_r>0$
ignored point on the right-hand side. That is,
\begin{equation}\label{eq:defk}
    k\in[-N_l,\,-w_l]\cup[w_r,\,N_r]
\end{equation}
We optimized those
parameters to get the best signal-to-noise ratio,
while maintaining the number of total
points to be equal to the ``SavGol'' case.

The third procedure avoids the bias by introducing
a blinded region around the peak altogether. 
By construction, this filtering method
allows for a perfect peak reconstruction
(Figure \ref{fig:Legend} and \ref{fig:ALor}, blue).
This means to disregard all the bins within a
certain window $k\in[0,\,k_{\mathrm{sig}})$, large
enough to cover the most of the signal---we recall 
that, because of \eqref{eq:defSi}, $i=0$ marks 
precisely the start of the signal window.

This method will be denoted as ``Blind.'' It is
dependent on the analyzed frequency, in that a 
different window is required for any positions of
the peak, built in such a manner to exclude the peak
region. In a real analysis, this means to run the
filtering algorithm for each tested hypothesis.
In other words, sliding a \emph{set} of windows
for each tested hypothesis, rather than sliding a
single window for every hypothesis. Numerically, 
the task is largely facilitated by the
construction of look-up tables. From the point of
view of the code, the case is similar to the
``Pierced'' one, but with the four parameters
$w_l$ and $w_r$ left free to assume negative
values. This makes sense, if we want the hole to be between $-N_l$ and $0$ or $0$ and $N_r$---see
\eqref{eq:defk}.
In the following, we will assume
a blinded region of $N_{bl} = 250$ bins, while the 
filtering window is symmetrical with total length
$N_p+N_{bl} = 1451$.

However, the amount of peak reduction in itself is 
not a good indication of the efficiency of the
method. Up to this point we have focused just on
the shape of the signal, disregarding any
fluctuations and/or their correlations.
As we will discuss in the next section,
one should also consider the
enhancement or suppression of the scale of the
fluctuations which can further reduce or highlight
any possible signal.

As we will show below, the standard SG
filtering is characterized by the highest
fluctuation suppression. For this reason, we
introduce the last method, which aims to combine
the unbiased results of the ``Blind'' method and
the minimum correlation given by the ``SavGol''
one. In this procedure, denoted as ``Double Fit''
(or ``DoubFit''), the ``Blind'' procedure is used to
get the filtered point just in the blinded window.
Then, a standard SG filtering is
performed, with the data in the signal region
replaced by the output of the ``Blind'' procedure.

%%%%%%%%%%%%%%%%%%%%%%%%%%%%%%%%%%%%%%%%%%%%%%%%%%%%
\section{Peak efficiency reconstruction}
\label{sec:effRec}
%%%%%%%%%%%%%%%%%%%%%%%%%%%%%%%%%%%%%%%%%%%%%%%%%%%%
The scope of this paper is not the
reconstruction of actual physical quantities but
rather the comparison of different algorithm.
Therefore, in order to quantify the problem, we
follow the approach by \cite{Yi:2023ekw}.

Specifically, we will tackle the problem of the
reconstruction of a peak above a flat baseline. In
other words,  without the features of the cavity
profile. that is, $b_i\equiv 0$. Moreover, we will
focus on the signal $s_i$ as reproduced by the
following toy parameterization, similar in shape to
the boosted Maxwellian expected for axions:
\begin{equation}
    \label{eq:defSi}
    s_i = 
    \begin{cases}
        \delta_a e^{-0.03\,i}\sinh(\sqrt{0.03\,i})
        & \text{for } i\geq 0;\\
        0 & \text{for } i < 0;\\
    \end{cases}
\end{equation}
where $\delta_a$ is a normalization constant
chosen according to our purposes. Because of its
pure mathematical nature, we can simplify the
problem even further by considering unitary
variance for the fluctuated data, i.e.\ $y_i\sim
\mathcal{N}(s_i, 1)$.

\begin{table}[t]
\begin{adjustbox}{max width=\textwidth}
\begin{tabular}{lcccc%
r@{\hskip0.15em}c@{\hskip 0.15em}l
r@{\hskip0.15em}c@{\hskip 0.15em}l
cr@{\hskip0.15em}c@{\hskip 0.15em}l}
\toprule
A & $N_l$ & $w_l$ & $w_r$ & $N_r$
& \multicolumn{3}{c}{$\xi'$}
& \multicolumn{3}{c}{$\xi$}
& $\delta_\mathrm{out}/\delta_\mathrm{in}$
& \multicolumn{3}{c}{%
$\varepsilon_{\text{\scshape{snr}}}$}\\
\midrule
True    &--- &--- &--- &---
        & 0.999 &$\pm$ & 0.045
        & 1.001 &$\pm$ & 0.002
        & 1.000
        & 0.999 &$\pm$ & 0.002\\
SavGol  &600 &--- &--- &600
        & 0.999 &$\pm$ & 0.045
        & 0.817 &$\pm$ & 0.001
        & 0.677
        & 0.829 &$\pm$ & 0.001\\
Pierced &675 &75 &175 &775
        & 1.000 &$\pm$ & 0.045
        & 1.161 &$\pm$ & 0.002
        & 0.947
        & 0.816 &$\pm$ & 0.001\\
Blind   &725 &--- &--- &775
        & 0.999 &$\pm$ & 0.045
        & 1.240 &$\pm$ & 0.002
        & 0.999
        & 0.806 &$\pm$ & 0.001\\
DoubFit &600 &--- &--- &600
        & 0.999 &$\pm$ & 0.045
        & 1.006 &$\pm$ & 0.002
        & 0.827
        & 0.822 &$\pm$ & 0.001\\
\bottomrule
\end{tabular}
\end{adjustbox}
\caption{Parameters $\xi'$, $\xi$,
$\delta_\mathrm{out}/\delta_\mathrm{in}$, and 
$\varepsilon_{\text{\scshape{snr}}}$ computed for
the filtering methods discussed in Section
\protect\ref{sec:FiltMeth}.}\label{tab:Eff}
\end{table}

The efficiency in the peak reconstruction,
$\varepsilon_{\text{\scshape{snr}}}$, can be
estimated by comparing the signal to noise ratios
in input and output. Following definition of Ref.\ 
\cite{Yi:2023ekw} (to which we address the reader
for details) we consider:
\begin{equation}
    \varepsilon_{\text{\scshape{snr}}} =
    \frac{1}{\xi}
    \frac{\mathrm{SNR}_{\mathrm{out}}}%
    {\mathrm{SNR}_{\mathrm{in}}}\simeq\frac{1}{\xi}
    \frac{\delta_\mathrm{out}}{\delta_\mathrm{in}}.
\end{equation}
The ratio $\delta_\mathrm{out}/\delta_\mathrm{in}$
can be thought as an averaged sum of the estimated 
signal \eqref{eq:hatSi} over the unfiltered signal
$s_{k}$ over some window 
$k\in[0\,k_{\mathrm{sig}})$:
\begin{equation}\label{eq:defEsig}
    %\varepsilon_{\text{\scshape{snr}}}
    \frac{\delta_\mathrm{out}}{\delta_\mathrm{in}}=
    \frac{\sum_{k=0}^{k_{\mathrm{sig}}-1}
    s_k\,\hat{s}_k}{\sum_{k=0}^{k_{\mathrm{sig}}-1}
    s_k^2}.
\end{equation}
Here, $s_k$
% in equation \eqref{eq:defEsig}
is given by \eqref{eq:defSi} and $\hat{s}_k$ by
\eqref{eq:hatSi}, with $b_k\equiv 0$. In our case,
we will consider $k_{\mathrm{sig}} = 250$, while
the parameters of the different filtering methods
can be found in Table \ref{tab:Eff}.

The factor $\xi$, on the other hand, is a 
normalizing factor that compensates for the change 
in the scale of the fluctuations.
Those arise from the bin-to-bin correlations caused
by the filtering methods. The  factor $\xi$ can be
though as the standard  deviation of $\hat{s}_k^
{\mathrm{add}}$, a weighted sum of $\hat{s}_k$ 
within the signal window. That is:
\begin{equation}
    \hat{s}_k^{\mathrm{add}} =
    \frac{\sum_{k=0}^{k_{\mathrm{sig}}-1}s_k\,
    \hat{s}_k'}{\sqrt{\sum_{k=0}
    ^{k_{\mathrm{sig}}-1} s_k^2}}.
\end{equation}
In this case, however, $\hat{s}_k'$ is built from 
the fluctuations of $y_k$ when $s_k = 0$; i.e.,
$y_i\sim\mathcal{N}(b_i,\,1)=\mathcal{N}(0,\,1)$.
That is, in our notation:
\begin{equation}
    \hat{s}_k' =
    (y_k - \hat{b}_k)/\xi'.
\end{equation}
Once again, $\xi'$ preserves the magnitude of the
standard deviation from $y_k$ to $\hat{s}_k'$ in
the signal window.
The value of the estimated parameters are reported
in Table \ref{tab:Eff}, for the different filtering
methods presented in Section \ref{sec:FiltMeth}. In
our case, we estimated $\xi'$, $\xi$ by means of
Monte Carlo evaluation on \num{2e5} samples.

Despite the SG filtering being the
worse at recovering the height of the peak, the
proposed alternatives get penalized by the
correlations among bins, canceling out any potential
gain.
%%%%%%%%%%%%%%%%%%%%%%%%%%%%%%%%%%%%%%%%%%%%%%%%%%%%
\subsection{Monte Carlo approach
to the signal recovery}
%%%%%%%%%%%%%%%%%%%%%%%%%%%%%%%%%%%%%%%%%%%%%%%%%%%%
We can test the robustness of these conclusion by
means of a simulated experiment. That is, a Monte 
Carlo-based likelihood approach built on the
formalism presented in Ref.\ 
\cite{GalloRosso:2022mhx}. 

\begin{figure}[t]
    \centering
    \includegraphics[width=0.75\textwidth,%
    valign=t]%
    {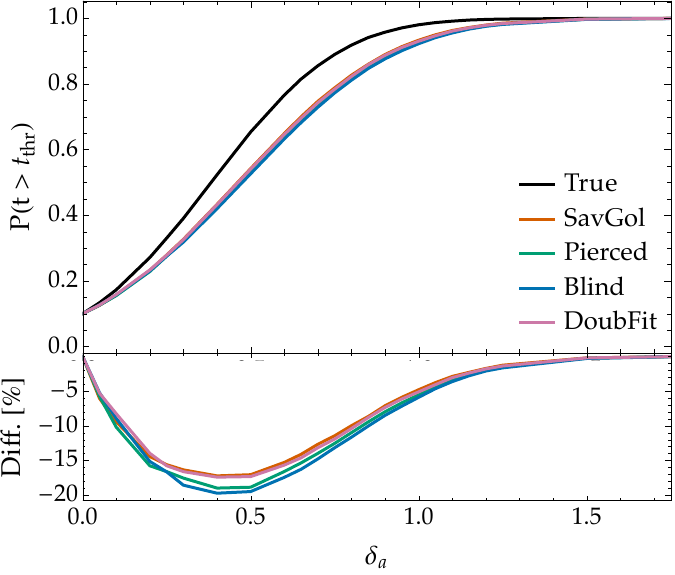}
	\caption{Top: probability of the claim of
    a discovery, as a function of $\delta_a$,
    for different filtering methods, with CL = 90\%.
    Bottom: deviation of the different filtering
    methods with respected to the perfect baseline
    reconstruction (Monte Carlo truth, represented by the black solid line in the top plot).}
	\label{fig:rMC}
\end{figure}

In this setting, we assume a value for the coupling
$\delta_a$. Then, pseudo-random data $y_i\sim
\mathcal{N}(s_i,\,1)$ are generated. They are then
fed into the filtering algorithm in order to
recover the reconstructed baseline. The noise 
fluctuations are quantified according to their
deviation with respect to the reconstructed 
baseline. This is similar to what would happen in a 
real analysis. Then, the test statistics $t$ is
computed,
\begin{equation}
    t=\frac{\sum_i s_i\,y_i}{\sqrt{\sum_i s_i^2}}.
\end{equation}
The value of $t$ is compared against a threshold,
$t_{\mathrm{thr}}$, defined with respect to the
null hypothesis ($\delta_a = 0$):\footnote{Note 
that, in most axion analyses, the null hypothesis
is defined as the existence of an axion.}
\begin{equation}
    P(t > t_{\mathrm{thr}}\,|\delta_a =0) =
    1 - CL,
\end{equation}
where CL is the confidence level, which we assumed
to be 90\%. If 
$t>t_{\mathrm{thr}}$, we claim that the null
hypothesis has been rejected and that our set of 
data allows for the claim of a ``discovery''.

We repeat the procedure for different values of
$\delta_a$, for different filtering algorithms,
registering the fraction of successful experiments.
This becomes our definition of efficiency.
Note that $t_{\mathrm{thr}}$ is built from simulated
data, too. Thus it automatically accounts for
the correlations among bins, induced by the
filtering procedures.

The values of $P(t>t_{\mathrm{thr}})$ are shown in
Figure \ref{fig:rMC} for the different filtering
methods. By construction, all the methods converge
at $1-CL = 0.1$ for $\delta_a = 0$. For coupling
$\delta_a$ strong enough, the signal becomes so
strong that the null hypothesis is rejected with
probability 1, regardless of the method used to
recover the baseline. In between, we see that
the performance of the filtering methods are all
within a few percent, and under-perform with respect
to the Monte Carlo truth (perfect knowledge of the
baseline) up to $20\%$, in line with the 
outcome of Table \ref{tab:Eff}.

\begin{figure}[t]
    \centering
    \includegraphics[width=0.75\textwidth,%
    valign=t]%
    {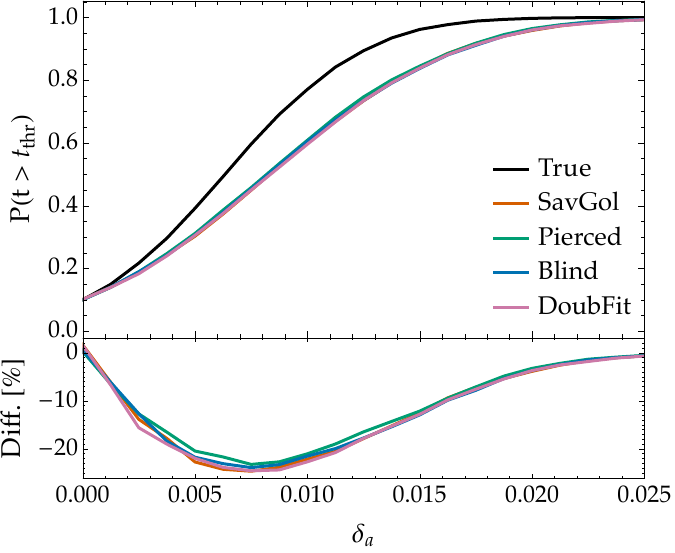}
	\caption{Same as Figure \ref{fig:rMC} with
    simulated data drawn from the baseline in Figure \ref{fig:bline}.}
	\label{fig:rMC2}
\end{figure}

The results are robust. In fact, the conclusions are
the same if we apply the filtering methods to a 
baseline more similar to a real one (Figure  \ref{%
fig:bline}, right). In this case, we attempted a 
reproduction of the standard aimed at a mature 
stage of the ALPHA experiment 
\cite{PhysRevD.107.055013}.
The baseline is assumed to have a
functional form as described by \cite{ADMX:2018gho},
with cryogenic quality factor $Q\sim\num{e5}$ and a
resonance of \SI{10}{\giga\hertz}.
%%%%%%%%%%%%%%%%%%%%%%%%%%%%%%%%%%%%%%%%%%%%%%%%%%%%
\section{Conclusions}
\label{sec:conc}
%%%%%%%%%%%%%%%%%%%%%%%%%%%%%%%%%%%%%%%%%%%%%%%%%%%%
In conclusion, in this paper we have applied 
different extensions
to the SG filtering to the problem of
the reconstruction of a signal $s_i$, expressed
as a peak over a slow-varying background $b_i$,
when only the combination $L_i=b_i+s_i$ is known,
and no description of $b_i$ is available. We have
applied the SG filtering, as well as
possible extensions, to compensate
for biased reconstructions and signal reduction.
We have quantified the efficiency using two 
different approach: the definition of 
\cite{Yi:2023ekw} and a Monte Carlo simulation.

Overall, the standard SG algorithm is
the one most biased in the peak reconstruction.
However, once the correlations among bins
are taken into account, all the methods are
characterized by the same efficiency in the 
reconstructed peak---within a few
percent---and with the standard SG
performing slightly better overall. For this reason,
axion-search experiments such as ALPHA would not
benefit from fancier extensions, more
complex and more computationally intensive overall.

\begin{figure}[t]
    \centering
    \includegraphics[width=0.75\textwidth,%
    valign=t]
    {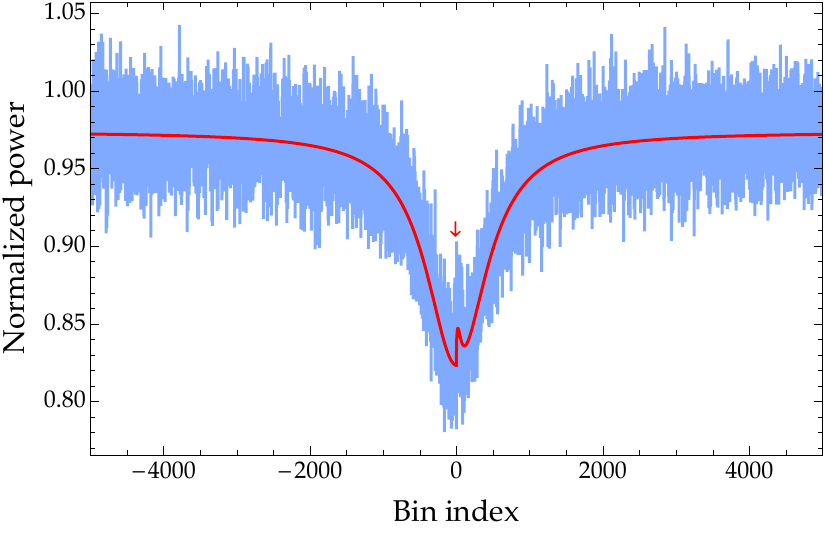}
	\caption{Simulated baselinew ith cryogenic 
    quality factor $Q\sim\num{e5}$ and a
    resonance of \SI{10}{\giga\hertz}.}
	\label{fig:bline}
\end{figure}

\section*{Acknowledgement}

The Authors acknowledge support from the Knut and 
Alice Wallenberg Foundation and Olle Engkvists 
Foundation.

\begin{appendices}

%%%%%%%%%%%%%%%%%%%%%%%%%%%%%%%%%%%%%%%%%%%%%%%%%%%%
\section{SG filter}
%%%%%%%%%%%%%%%%%%%%%%%%%%%%%%%%%%%%%%%%%%%%%%%%%%%%
\label{App:SavGol}
In this section, we report the calculation of the
coefficients $c_k$ appearing in \eqref{eq:defSGen},
for the reader's convenience. As the matter has
extensively been dealt with in the literature, we will
follow the reasoning as presented in Refs.\  
\cite{Schafer2011,NumRec2,NumRec3}.

Let us consider a vector of measured values $\bm{y}$,
corresponding to some independent variable $\bm{x}$.
In the case of the SG filtering, for instance, $\bm{x}$
is built to encompass all the points around the
filtered one. We are  interested in fitting the
measured values $\bm{y}$  with a fitting function
in polynomial form:
\begin{equation}
	\label{eq:FitFun}
    f(x) = \sum_{j = 0}^n a_{j}\, x^j.
\end{equation}

Note that the function is obtained as a fit around a
specific bin $x_i$, and that the set of coefficient $a_{j}$ depends on this particular choice through the
$N_p$ measured values that enter in the fit. However,
carrying the bin index $i$ would have made the notation
unnecessarily heavy.

The coefficients $a_{j}$ are chosen to minimize the
mean-squared approximation error for the group of
input samples:
\begin{equation}
    \mathcal{E} = \sum_{\ell=0}^{N_p-1}
    \left([\bm{f}]_\ell-[\bm{y}]_\ell\right)^2,
\end{equation}
where we have defined
\begin{equation}
    [\bm{f}]_\ell = f_\ell = f(x_\ell).
\end{equation}

We can simplify the problem if we assume to deal
with a series of equally-spaced points, as it 
happens for the applications discussed in this
paper. In such a case, the absolute scale of the
independent variable $x$ is not important, as
everything can get shifted and rescaled. In other
words, we can define the index $k$ as the (discrete)
distance from the filtered value (now, zero):
\begin{equation}
    \mathcal{E}_i=\sum_{k=-N_l}^{N_r}\left(\sum_{j=0}^n
    b_{j}\,k^j - y_{k}\right)^2.\footnote{We did not explicitely write $y_{k+N_l}$, as we thought that
    $y_k$ was clearer in highlighting that the boundaries have been properly shifted.}
\end{equation}
Differentiating by the $b_{j}$ and setting
the corresponding derivative equal to zero
yields the set of $n + 1$ equations in $n + 1$
unknowns:
\begin{equation}
    \label{eq:bijMat}
    \sum_{j=0}^n\sum_{k=-N_l}^{N_r}k^{j+\ell} b_{j}= 
    \sum_{k =-N_l}^{N_r} k^{\ell}
    y_{k}\quad\text{for}\quad\ell = 0,\dots,\,n.
\end{equation}

As the filtered value corresponds to the fit
function evaluated at $k = 0$, the problem reduces to
the calculation of just one coefficient, $[\bm{b}]_0$.

A clever way to deal with this problem is to express
Equation \eqref{eq:bijMat} in matrix form:
\begin{align}
    \label{eq:normal}
    \left(\bm{V}^T\cdot\bm{V}\right)\cdot \bm{b}
    &= \bm{V}^T\cdot\bm{y},\\
    \bm{b}&=\left(\bm{V}^T\cdot\bm{V}\right)^{-1}
    \cdot \bm{V}^T \cdot\bm{y}\equiv
    \bm{H}\cdot\bm{y},
    \label{eq:normal2}
\end{align}
 where $\bm{V}$ is known as Vandermonde's matrix:
\begin{equation}
    \label{eq:defA}
    \left[\bm{V}\right]_{kj} = k^j,
\end{equation}
with $\left[\bm{V}\right]_{00} = 1$. The index $j$
goes from $0$ to the polynomial degree $n$, while
each column needs to include the $N_p$ points $k$,
from $-N_l$ to $+N_r$.

To get the coefficients $c_k$ in \eqref{eq:defSGen}
we are helped by the following relations:
\begin{align}
    \label{eq:VVrel}
    \left[\bm{V}^T\cdot\bm{V}\right]_{Kj} &=
    \sum_{k}[\bm{V}]_{ki}
    [\bm{V}]_{kj} = \sum_{k} k^{i+j},\\
    \left[\bm{V}^T\cdot\bm{y}\right]_{kj} &=
    \sum_{k} V_{kj}y_k =
    \sum_{k} k^j y_k.
\end{align}

Now, recapping, for the SG filtering we are interested
in the point:
\begin{equation}
    f(0) = b_0 = \sum_{k=-N_l}^{N_r}c_k y_k,
\end{equation}
and ultimately in the set $\bm{c}$ of coefficients.
Those are easily obtained if we notice that $c_k$ is exactly the output $b_0$ when all the components of
$\bm{y}$ are null but the $k$-th one.
Thus, we can simply substitute $\bm{y}$ with
the unit vector $\bm{e}_k$:
\begin{equation}
    \label{eq:getWn}
    c_k = \left[\left(\bm{V}^T\cdot
    \bm{V}\right)^{-1}\cdot\left(\bm{V}^T
    \cdot\bm{e}_k\right)\right]_0 =
    \sum_{j = 0}^n \left[\left(\bm{V}^T\cdot
    \bm{V}\right)^{-1}\right]_{0j} k^j.
\end{equation}

From Equation \eqref{eq:getWn}, we see that just the
first row of matrix $(\bm{V}^T\cdot\bm{V})^{-1}$
is needed. Therefore, efficient computations can
be implemented for such a purpose---see e.g.\ Section 2.3 in \cite{NumRec2,NumRec3}.

\end{appendices}

\printbibliography

\end{document}